\documentstyle[epsfig,aps,prl,preprint]{revtex}
\draft

\begin{document}

%\twocolumn

\hsize\textwidth\columnwidth\hsize
\csname@twocolumnfalse\endcsname

\title{Transport properties of vortices in easy flow channels: a Frenkel-Kontorova study}

\author{R.~Besseling, R. Niggebrugge, and P.H.~Kes}

\address{Kamerlingh Onnes Laboratorium, Leiden
University, P.O. Box 9506, 2300 RA Leiden, the Netherlands}
\date{\today}
\maketitle

\begin{abstract}
Incommensurate easy flow channels in an otherwise perfect vortex
lattice are investigated. The associated (point) defects in the lattice inside the
channel cause an almost vanishing critical current, as shown by molecular dynamics
simulations and a comparison with the Frenkel-Kontorova model. In addition
to the normal flux flow behavior we find a low mobility regime at small
drives associated with defect motion. We treat this situation analytically for the case
of a single defective vortex row. We also briefly discuss the relation to existing
experiments on artificial vortex channels.
\end{abstract}
%\twocolumn
%\begin{multicols}{2}
%\raggedcolumns
\narrowtext
\noindent
In recent years, vast interest has grown in the dependence of static
and dynamic properties of
vortex matter on the spatial profile of the pinning potential originating
from defects in the underlying
material. The behavior of the vortex lattice was investigated experimentally
as well as theoretically in the
presence of a wide variety of these pinning landscapes, ranging from purely
random \cite{Koshelev,JensenPRL,JensenJLTP} to highly periodic \cite{Reichhardt}.
Several of these studies revealed the existence of static channels of easy vortex flow at
currents just above the critical value \cite{JensenPRL,JensenJLTP}. In addition,
this critical current was shown to be proportional to the interaction
strength between vortices inside the channel and those in the
channel edges (CE's), as expressed by the shear modulus of
the vortex lattice $c_{66}$ \cite{JensenJLTP,Pruijmboom}. A precise
theoretical description of this phenomenon and of its
dependence on the channel width is still lacking. In this paper
we study for the first time the properties of static easy flow
channels within the framework of the Frenkel-Kontorova (FK)
model \cite{Frenkel}. It is demonstrated that a mismatch
between channel width and lattice constants induces (point)defects
in the channel leading to an almost vanishing shear strength.
This can have important implications for the properties of
vortex matter in a pinning potential with large spatial variations
in strength. In addition, our results on the dynamics of incommensurate
structures may be applied to various other fields, such as vortex
dynamics in overdamped Josephson junction arrays, transport
properties of charge density waves \cite{Floria} and solid
friction of confined layers \cite{BraunPRB97,Kawaguchi}.

We consider 2D vortices at $T=0$ in an easy flow channel of
width $w$ and length $l$, confined by two semi infinite pieces
of a rigidly pinned vortex lattice with perfect hexagonal structure
of lattice constant $a_0$ and row spacing $b_0=a_0\sqrt{3}/2$.
The pinned structures forming the edges are chosen to have their
principal axis in the $x$-direction parallel to the channel and
have a relative shift $\Delta x$ (left upper corner of Fig. 1a). First we describe
qualitatively the case in which $w \approx b_0$ and $\Delta x=0$
(restricting the degrees of freedom for the mobile vortices to the
$x$ direction), corresponding to the single chain FK model.
Molecular dynamics (MD) simulations will then provide a
justification of these qualitative arguments as well as show that
our line of reasoning also applies to larger channel widths and
a relative CE-shift $\Delta x\neq0$.

When a shear force is applied to the vortices in the channel, the
onset of flux flow appears as the force exceeds the maximum
shear stress. The maximum shear force density $F_s$ may be
obtained from $c_{66}$: $F_s=2Ac_{66}/w$ where $A$
accounts for the lattice anharmonicity and orientation of
the shear direction with respect to the principal lattice
axes \cite{Frenk2}. For our orientation and a harmonic
potential from the vortices in the CE's, it follows that
$A=(\pi\sqrt{3})^{-1}$. Then, for $w \approx b_0$, the
'edge' potential (per unit length) for a single vortex inside
the channel can be defined as $U=\frac{\sigma}{2} (1+\cos(k_0x))$,
with $\sigma={a_0}^3 c_{66}/(\pi ^2 b_0)$ and $k_0=2\pi/a_0$.
Next, we consider the mutual interaction between the mobile
vortices in the channel. These vortices have an average spacing
$a$ which is given by the induction B as $a=\Phi_0/(Bw)=a_0b_0/w$
with $\Phi_0$ the fluxquantum. We use a harmonic approximation for
the interaction, while its nonlocal nature is preserved by including
interactions with $n$ neighbors in the chain. The effective spring
constants $K_n$ then can be obtained from a London $1/r$ interaction
force between vortices: $K_n= (\Phi_0^2/4\pi\mu_0\lambda^2)(1/na)^2$
with $\lambda$ the penetration depth. Thus, in the presence of a transport
current density $J$ applied perpendicular to the channel , the equation
of motion for vortex $i$ in the channel reduces to an overdamped FK model:
\begin{equation}
\gamma \dot{x_i}=f+\mu\sin(k_0x_i)+\sum_{n}K_n(x_{i+n}+x_{i-n}-2x_i),
\label{eqmot}
\end{equation}
\noindent
where $f=J\Phi_0$ is the Lorentz force, $\mu=\pi\sigma/a_0$ is the
amplitude of the periodic force, and the friction coefficient $\gamma$ is related
to the flux flow resistance by $\gamma=B\Phi_0/{\rho_{ff}(B)}$.

In the FK-model the critical force $f_c$ and the dynamic properties largely
depend on both the value of $a/a_0$ as well as the dimensionless elastic
constant $g=a_0^2\sum_{n}n^2K_n/2\pi^2\sigma$ \cite{Floria,BraunPRB94}.
From the above given expressions for $\sigma$ and $K_n$ and using the low field
expression for the shear modulus $c_{66}=\Phi_0B/(16\pi\mu_0\lambda^2)$,
we find that $g \approx (\lambda/a)(a_0/a)^2$ for our system. Since in practice
\cite{Pruijmboom} $\lambda/a_0>1$ we restrict ourselves to the case $g>1$
\cite{footnote}. In case $a/a_0=1$, the vortex row locks in with the periodic potential
and $f_c=\mu$. For rational $a/a_0$, the row contains point defects
with a size $l_d\approx 2\pi a\sqrt{g}$. Then $f_c$ is determined by their Peierls-Nabarro
(PN) barrier \cite{BraunPRB94} which, for an isolated defect and $g>1$ is given
by $f_c\approx (8/3)\pi^3g\mu e^{-\pi^2\sqrt {g}}$.
Accordingly, $f_{c}/\mu\ll1$ and the channel has an almost vanishing
critical current. Finally, for irrational $a/a_0$ and $n=1$, Aubry has shown
that a transition exists to a state with $f_c=0$ as $g$ exceeds a critical
value $g_c$. For $a/a_0=(\sqrt{5}-1)/2$, $g_c=1$ \cite{Aubry}. The condition
$g>1$ then implies that the vortex channel can have a truly vanishing critical current.
The influence of nonlocal elastic constants \cite{Brandt} on the above results can be neglected as long
as $l_d\gtrsim\lambda$, i.e. for $\lambda\lesssim 50a_0$.

In order to check the above scenario, we performed MD
simulations of easy flow channels in a thin film. The dynamics of
a vortex inside the channel is governed by the following equation:
\begin{equation}
{\dot{\bf r} _i} = f +\sum_{j\neq i} {\bf f}_v({\bf r}_{ij}) ,
\label{moldyn}
\end{equation}
with $a_0$ as unit of length, $\Phi_0^2/(2\pi\mu_0\lambda^2a_0)$
as unit of force (per unit vortex length) and $4\pi\mu_0\lambda^2/(\sqrt{3}\rho_{ff})$
as unit of time. The external drive $ f$ is directed along the channel.
The interaction force is chosen as $f_v = (1/r)(1-r^2/r_c^2)^2$ with
$r_c=3.33$ \cite{Koshelev}. This interaction results in a
shear modulus coinciding with the above given expression (in dimensionless
units $c_{66}=1/(4\sqrt{3}))$. In Eq. (\ref{moldyn}), the index $j$ runs over
all vortices, including those in the CE's (see Fig. 1a). We adopted
periodic boundary conditions with $l$ large enough, such that our
results did not depend on this length scale. For each $w$, we
relaxed the system to its ground state (GS) and then measured
the force-velocity ({\it f-v}) characteristic by stepwise cycling the force
($f=0\rightarrow f_{max} \rightarrow 0$) and taking the
stationary velocity $v=\langle{\dot{x}}\rangle_{i,t}$ at each step.

Fig. 1a shows the resulting {\it f-v} curves for $\Delta x=0$ and
$w \approx b_0$. For $w=b_0$, the maximum shear strength is
observed with a value $f_c=0.054$, coinciding with the dimensionless
value of $\mu$ in (\ref{eqmot}). Additionally,
the dynamics are identical to that of a single particle in a sinusoidal
potential. Both observations are in agreement with the fact that the
'edge' potential resulting from our vortex interaction is indeed harmonic.
More interesting is the behavior in the incommensurate
case ($w\neq b_0$). As seen in the inset $f_c/\mu \lesssim 0.001$, which
can be understood from our previous arguments when we calculate $g$:
expanding $f_v$ in (\ref{moldyn}) we obtain $K_n=(na/a_0)^{-2}+
0.18-0.003(na/a_0)^2$, resulting in $g\approx6$. A typical example of the
GS defect structure is shown in Fig. 1b, where the open symbols represent
the vortex density along the channel for $w=0.97b_0$.

Turning to the dynamics in Fig. 1a, it is seen that the defective vortex
rows posses a regime of low mobility in the {\it f-v} curves. In this regime,
transport in the row is carried by the defects which have a velocity
$v_d$ much larger than the average velocity $v=c_dv_d$, with
$c_d=|1-(a_0/a)|$ the defect density. A snapshot of the moving row then
resembles the GS defect structure (see again the open symbols in Fig. 1b).
A useful quantity here is the mobility of an isolated defect. From the simulation
we get $v_d/f\approx13$, a value which one can also obtain from the analytical
treatment described below. The mobility of an interstitial is slightly larger than
that of a vacancy due to the anharmonic vortex interactions. When entering the high
mobility regime where $f\gtrsim \mu$, the defects are smeared out in a slight sinusoidal
density modulation and the transport becomes coherent within the row (see the filled
symbols in Fig. 1b).

It turns out that we can capture the transport characteristics
within a perturbation theory as used in Ref. \cite{Strunz}.
Consider Eq. (\ref{eqmot}). It is known that the motion of all
particles in the FK chain can be completely described by the dynamic hull function
$h(ia+vt)=x_i(t)- (ia+vt)=h(\varphi)$, which represents the deviations of the
particle positions from the undisturbed sliding values.
From (\ref{eqmot}) we obtain the following equation for $h(\varphi)$:
\begin{eqnarray}
{\gamma}v(1&&+h'(\varphi)) =f+\mu\sin(k_0(\varphi+h(\varphi)))\nonumber\\
&&+\sum_{n}K_n(h(\varphi+na)+h(\varphi-na)-2h(\varphi)).
\label{hulleq}
\end{eqnarray}
When mapping $\varphi$ back onto one period of the
potential, $\tilde{\varphi}\equiv(\varphi-ia_0)$,
$h(\tilde{\varphi})$ has the periodicity $a_0$ resulting in:
\begin{eqnarray}
{f}={\gamma} v 
+ \frac{\gamma v}{a_0} \int ^{a_0}_{0}
[h'(\tilde{\varphi})]^2 d \tilde{\varphi}.
\label{Fv}
\end{eqnarray}
The shape and amplitude of $h(\tilde{\varphi})$ represent the defect
structure of the moving FK chain and thus depend on $a/a_0$, $g$ and $v$.
For large speed or in case of both large lattice mismatch and $g\gg 1$ (above
Aubry's transition), $h(\tilde{\varphi})$ is sinusoidal with a small amplitude \cite{Kawaguchi}.
In our case, $h(\tilde{\varphi})$ can have a more complex shape as well as
a large amplitude. We introduce higher orders in the perturbation to describe
the dynamics for any $a/a_0$ by solving (\ref{hulleq}) with a trial hull function of
the form $h(\tilde{\varphi})= \sum^M_{m=1}\left(H_m e^{ik_0m \tilde{\varphi}}+ c.c.\right)$.
Inserting this hull function in (\ref{hulleq}), we obtain a set of coupled equations for
the coefficients $H_m$. Solving for $H_m$ yields the approximate form of $h(\tilde{\varphi})$.
A full account of this work will be published elsewhere. Now the expression describing
the {\it f-v} characteristics is obtained by using this hull function in Eq. (\ref{Fv}), which
results for $M=3$ in:
\begin{eqnarray}
f=\gamma v +\frac {2\pi^2\mu^2{\gamma}v}{a_0^2\Omega_{c}^4+(2\pi{\gamma}v)^2},
\label{ptbeq}
\end{eqnarray}
where $a_0\Omega_c^2$ represents the elastic restoring force on {\it nonlinear}
excitations in the FK chain. From the perturbation approach $\Omega_c$ is
expressed in $\mu$ and the restoring forces on {\it linear} modes (phonons)
at wavenumber $k=mk_0$:
\begin{eqnarray}
\Omega_c^2(a=a_0b_0/w)=\frac{\Omega_1^2\Omega_2^2\Omega_3^2a_0^2+\mu^2\pi^2(\Omega_
1^2+\Omega_3^2)}
{\Omega_2^2 \Omega_3^2a_0^2+\mu^2\pi^2 },
\label{nldisp}
\end{eqnarray}
in which $\Omega_m^2=\Omega^2(mk_0)=\sum_n2K_n(1-\cos(nmak_0))$.
The last term in Eq. (\ref{ptbeq}) corresponds to the sliding friction in the field of
tribology \cite{Kawaguchi}. It results from the dissipation of the internal (nonlinear)
collective modes in the chain. The crossover from low to high mobility occurs
at $v_c=a_0\Omega_c^2/(2\pi\gamma)$ where the amplitude of the hull function
drastically decreases. In order to compare
(\ref{ptbeq}) with the simulated data, we use the previously obtained expression
for $K_n$ and take into account that $\mu$ slightly depends on $w$ as well.
The corresponding {\it f-v} curves, shown in Fig.1a, agree well with
the simulated data. The transport of a defective vortex row can thus be
accurately described by perturbing the harmonic FK model.

In order to study channels of larger width, we adopted an edge shift $\Delta x(w)$
with a saw tooth shape ($0 \leq \Delta x \leq a_0/2$) which assures that, as
we vary $w$, a perfect hexagonal structure is retained for $w=pb_0$
with $p$ an integer. However, for $w\neq pb_0$, the qualitative
behavior did not depend on $\Delta x$. The inset to Fig. 2 shows
the simulated {\it f-v} curves for $3.5b_0<w<5b_0$. The features
are similar to the {\it f-v} curves for $w\approx b_0$ and these are
again associated with the presence of defects in the channel.
Two characteristic structures can be identified: for
$\mid(w/b_0)-p\mid\lesssim0.4$ (I) the structure consists of
$p$ vortex rows plus interstitials ($w/b_0>p$) or vacancies
($w/b_0<p$). The defects are not necessarily equally distributed
within a row. In addition the defect density differs per row due to the repulsive
interaction with the channel edges. The latter is illustrated in Fig. 3a, where two
snapshots of the motion for $w=4.08b_0$ in the low mobility
regime are displayed with interstitials indicated by arrows.
For $\mid(w/b_0)-(p+1/2)\mid\lesssim0.1$ (II), clustering of
vacancies ($w/b_0>p+1/2$) or interstitials ($w/b_0<p+1/2$)
occurs. As a result, alternating regimes form along the channel
of $p$ and $p+1$ rows, which we refer to as stacking faults (SF's).
The regimes are separated by dislocations with Burgers
vectors at angles of $60^{\circ}$ with the channel direction, as
shown in Fig. 3b.

Structure I is just a $2D$ extension of our results for a single incommensurate
vortex row. The shear strength is almost vanishing due to the small PN barriers
of the point defects in each row. The low mobility regimes in
the {\it f-v} curves once more originate from motion of these defects. However,
two additional features appear. First, recalling our simple description of defect
motion in a row ($v_r=c_{d,r}v_d$), we see that, for equal defect velocities in
different rows, a different defect density causes different average row
velocities $v_r$, i.e. {\em plastic motion within the channel}. This clearly shows up
in Fig. 3a, where the interstitials move at the same velocity $v_d \gg v_r$ while the
row velocities vary from $v_2=0$ to $v_3=2v$. Second, occasionally a hysteresis
is observed in the transport curves of I (see inset to Fig. 2). This hysteresis originates
from either a redistribution of defects {\it between} rows (for $0.2\lesssim|w/b_0-
p|\lesssim 0.4$) or {\it within} the rows. After the transition, defects within the row have
the maximum possible separation and they are clustered with those in neighboring
rows, as expected from the standard FK model. Now the mobility of a defect
moving along a locked (defect free) region of a neighboring row is small
compared to that of a defect traveling along with a neighboring defect,
since part of the periodic potential is destroyed in the latter case.
This explains the transport hysteresis and the fact that the mobility is always larger
after the transition. However, the nature of the transition and whether or not it
occurs depend both on $\Delta x$ and $w$. In the high mobility regime of I,
vortex transport is no longer carried by the defects but by the rows as a whole.
They attain the same velocity and the flow loses its plastic character.

The SF's of II can locally form a commensurate structure. Then one
naively expects $f_c$ to be of the same order as the commensurate value.
The actual value, $f_c \approx \mu b_0/(15w)$, is much lower because the SF's
can move plastically by continuously creating and annihilating the dislocations at
their borders. This is supported by the neighboring, defective rows:
these rows 'feed' the SF with point defects so that the longitudinal
vortex displacements change into transverse displacements when
entering the SF, as shown by the trajectories in Fig 3b. Now the naive
estimate, $f_c\approx \mu b_0/(2w)$, marks the crossover
to the regime where the SF's move coherently with the rest of the structure:
$v_{SF}=v$. The hysteresis, observed in the {\it f-v} curve for $w=4.54b_0$, is now
caused by transitions in which the number of SF's decreases. For $f\downarrow$, this
appears as a reduction in the amount of plastically moving SF's and thus an increase in mobility.

The main plot of Fig. 2 summarizes the behavior of $f_c$ vs. $w$,
where we took the maximum $f_c$ in case of hysteresis. The continuum
result $f_c= \mu b_0/w$ applies only for $w=pb_0$ while $f_c$
of the incommensurate channels reflects our above discussion of the
barriers of the corresponding defects. To make a link to the experiments
on samples with artificial vortex channels as used in Ref. \cite{Pruijmboom}, we
also show the measured shear strength $f_c=2\pi J_c\mu_0 \lambda^2a_0/ \Phi_0$
versus $w/b_0$ of such a device. It is seen that the commensurability peaks are
lower and smeared out. Preliminary results of simulations in which positional
disorder of vortices in the CE's is implemented, indicate that pinning of defects
(at $w\neq pb_0$) and generation of defects (at $w=p b_0$) by this (phase)
disorder in the periodic potential forms the mechanism for the smoothening of $f_c$.

In conclusion, we have shown that the critical current of incommensurate vortex
flow channels is drastically reduced as compared to the expected value
for commensurate channels. The dynamic behavior exhibited a crossover
from defect motion to coherent flow. We described these transport characteristics
analytically for the case of a single vortex row.

We would like to thank A.E. Koshelev, A. van Otterlo and T. Droese
for stimulating discussions.

%\end{multicols}

\begin{figure}
\label{I}
\vspace{0cm}
\caption{(a) {\it f-v} characteristics for vortex channels with
$w \approx b_0$ and $\Delta x=0$. Symbols are simulated
data and drawn lines are obtained with Eq. (\ref{ptbeq}).
The inset shows an expanded view of the small velocity regime.
Left upper corner shows the channel geometry with mobile
vortices displayed by open symbols.
(b) Vortex density $\rho(x)=1/(w(x_{i+1}-x_{i}))$
along the channel for $w=0.97b_0$: ($\circ$) GS as well as a snapshot
of the moving row in the low mobility regime ($f=0.01$), ($\bullet$) snapshot
at large drive ($f=0.1$). The drawn line represents
$\rho(x)=(w(h(\varphi+a)-h(\varphi)+a))^{-1}$ at $v(f=0.01)$.}
\end{figure}

\begin{figure}
\label{II}
\vspace{0.1cm}
\caption{Shear strength $f_c$ versus channel width for
a velocity criterion of $5 \times 10^{-4}$ and
$\Delta x=\Delta x(w)$. Open symbols are experimental data of
artificial vortex channels at the same criterion \protect\cite{Pruijmboom}.
The dashed line represents the continuum result: $f_c=\mu b_0/w$.
The inset displays typical simulated {\it f-v} characteristics. Note the hysteresis
for $w= 4.54b_0$ and $w=3.76b_0$.}
\end{figure}

\begin{figure}
\label{III}
\vspace{0.1cm}
\widetext
\caption{(a) Two snapshots of the channel flow for $w=4.08b_0$ (case I)
in the low mobility regime (f=0.007, see inset to Fig. 2). For clarity
the amplitude of vortex displacements from the average $y$-coordinate of
a row has been enlarged by a factor $3.5$. Snapshot $2$ ($\circ$) is
taken after an average displacement of $0.1a_0$ with respect to snapshot
$1$ ($\bullet$), whereas the defect displacement is about $6a_0$.
The rows are labeled with their interstitial density and average velocity.
Defects are indicated by arrows. (b) ($\circ$) defect structure at $w/b_0=4.54$ (case II),
small dots represent the trajectories in the plastic regime at $f=0.002$.}
\end{figure}

%\clearpage
%\newpage
%\ FIGURE CAPTIONS
\end{document}